# Hybrid deep learning and physics-based neural network for programmable illumination computational microscopy

Ruiqing Sun,[a] Delong Yang,[a] Shaohui Zhang,[a,*] Qun Hao[a,b,c]

[a] School of Optics and Photonics, Beijing Institute of Technology, Beijing 100081, China
[b] Changchun University of Science and Technology, Changchun 130022, China
[c] e-mail: qhao@bit.edu.cn

**Abstract**. Relying on either deep models or physical models are two mainstream approaches for solving inverse sample reconstruction problems in programmable illumination computational microscopy. Solutions based on physical models possess strong generalization capabilities while struggling with global optimization of inverse problems due to a lack of insufficient physical constraints. In contrast, deep learning methods have strong problem-solving abilities, but their generalization ability is often questioned because of the unclear physical principles. Besides, conventional deep models are difficult to apply to some specific scenes because of the difficulty in acquiring high-quality training data and their limited capacity to generalize across different scenarios. In this paper, to combine the advantages of deep models and physical models together, we propose a hybrid framework consisting of three sub-neural networks (two deep learning networks and one physics-based network). We first obtain a result with rich semantic information through a light deep learning neural network and then use it as the initial value of the physical network to make its output comply with physical process constraints. These two results are then used as the input of a fusion deep learning neural work which utilizes the paired features between the reconstruction results of two different models to further enhance imaging quality. The final result integrates the advantages of both deep models and physical models and can quickly solve the computational reconstruction inverse problem in programmable illumination computational microscopy and achieve better results. We verified the feasibility and effectiveness of the proposed hybrid framework with theoretical analysis and actual experiments on resolution targets and biological samples.

**\*** Shaohui Zhang**,** E-mail: zhangshaohui@bit.edu.cn

## 1 Introduction

This document shows the suggested format and appearance of a manuscript prepared for SPIE journals. Accepted papers will be professionally typeset. This template is intended to be a tool to improve manuscript clarity for the reviewers. The final layout of the typeset paper will not match this template layout.

Fourier Ptychographic microscopy (FPM) is a typical computational microscopy for its characteristics of optimizing the hardware and algorithms simultaneously for advanced better imaging results [1]. It breaks through the limits of traditional optical systems, offering high-resolution imaging over a wide field-of-view (FOV) even with low numerical aperture (NA) lenses [2,3]. The affordability of FPM, combined with its elimination of phase measurement needs, has made it particularly attractive in fields like digital pathology, surface analysis, and stem cell research. [4-7, 39]. Conventional high synthetic bandwidth-product (SBP) imaging typically depends on lenses with high NAs, necessitating detailed spatial scanning [7,9]. This approach is less suitable for in vivo imaging and often faces challenges in maintaining focus during the mechanical scanning process due to the lens's limited depth of field [7,8]. In stark contrast, FPM systems eschew complex mechanical structures in favor of a more streamlined approach. It transforms a standard microscope by adding a programmable LED array, enabling the capture of low-resolution images from various illumination angles. This process involves scanning the specimen in the frequency domain and iteratively refining the estimation of the sample's complex amplitude based on the captured intensities. Ultimately, it achieves high SBP imaging with an expanded depth of field [1].

However, the original FPM method still has some shortcomings. It entails sequentially taking low-resolution images from various illumination angles, with a requirement to maintain at least a 60% overlap in the Fourier space of the sample [9, 10]. This significantly slows down the imaging speed. Fortunately, researchers have found ways to overcome these problems by accurately modeling the physical processes of programmable illumination. Tian and colleagues first introduced a random multiplexing method to speed up the capture process [11]. They later extended their method by separating bright and dark fields [12], which further improved the image quality. However, their approach didn't fully consider how varying characteristics of specimens might impact the illumination model, leading to performance limitations. Additionally, these methods rely on finding analytical differentiation of imaging parameters, a process that's particularly complex for factors like the sample's defocus distance and variations in LED illumination intensity. Simultaneously optimizing multiple parameters also adds to the complexity and challenges of their methods. Kellman et al [13] developed an approach by creating a neural network that aligns with physical properties, tailoring the illumination model to samples with specific traits. This data-driven method allows for the automatic calculation of numerical differentiation and leverages training tools for optimal multi-parameter solutions, akin to training a neural network. However, it encounters obstacles in areas where establishing a gold standard is challenging, such as in diagnosing rare diseases. Additionally, its adaptability to different sample types is not always certain, potentially limiting its wider application. Sun et al. suggested a physics-based method to create adaptive illumination models for each sample without needing extra data [8]. While this method works well with 20 illumination patterns, its effectiveness might diminish when attempting to reduce the number of illumination patterns to as few as 10. The complexity of simultaneously using many LEDs poses a challenge for data analysis, making it difficult to obtain good results with existing optimization algorithms.

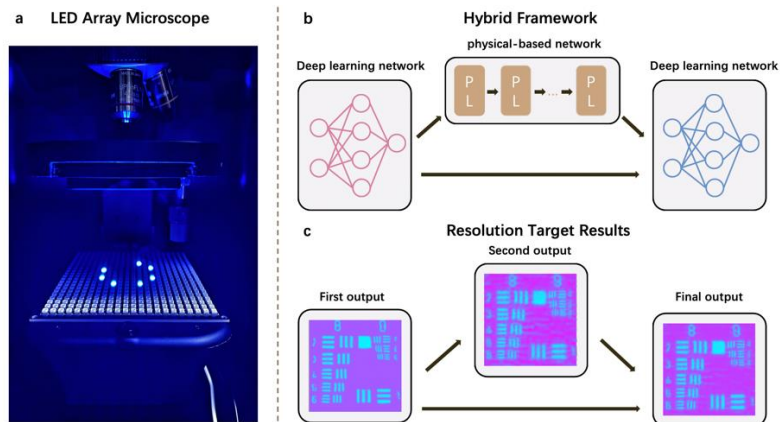

**Fig. 1** Here we show the imaging system we used and how our framework worked. (a) The system we used in experiments to verify the effectiveness of our framework. (b) The overview of the framework we proposed where PL is used to refer to the physical layer in the physical model. (c) Reconstruction results of 10 low-resolution resolution target images captured using a 0.13NA objective.

Contrasting with conventional methods that depend on detailed physics models, deep learning leverages data to learn and address inverse problems [21, 32]. This approach has notably impacted fields like natural language processing and computer vision [14-15, 33]. Additionally, its capability in tackling complex imaging challenges is gaining recognition, as indicated by recent research [16-21]. In computational microscopy, there are two primary deep learning approaches for solving inverse problems: 1. End-to-end solutions and 2. solutions based on Deep Image Prior (DIP) [23]. In the end-to-end approach, a neural network is trained to transform input images into the desired output for a specific microscopy reconstruction problem. Therefore, the closer the training set's distribution is to the experimentally collected data, the better the network performs in imaging. For instance, Nguyen et al. [22] trained their network using single-frame data from ongoing observations. This approach reduces the acquisition time of low-resolution images and speeds up sample reconstruction by decreasing the overlap rate requirements in Fourier space. Cheng and colleagues [24] took a slightly different approach if the sample fits a particular distribution. They achieved a high space-bandwidth-time product in FPM with only a single acquisition. However, these methods encounter challenges with rare samples that lack enough data for effective network training. Furthermore, if the actual sample being observed significantly deviates from the training data, this can lead to inaccurate imaging results. In contrast, DIP solutions integrate physics models and deep model assumptions to guide the update of network weights in a single imaging process, without requiring additional data. For instance, Situ et al. [21] created PhysenNet, a network tailored for single-beam phase imaging, utilizing the DIP method. Coincidentally, Zhang et al. [25] designed a network capable of reconstructing images from data characterized by a low rate of overlapping apertures in FPM. Their approach operates independently of external data, but it hinges on an accurate physics model and requires dynamic adjustments to network parameters for each image acquisition. These necessary adjustments can be both time-consuming and demanding in terms of resources.

In multiplexed FPM, it's difficult to get good results using only physical models, data-driven methods, or DIP methods, especially for schemes with a higher degree of programmable multiplexing illumination. However, our research achieved improved results by combining the advantages of both physical and deep learning models within a hybrid framework, as illustrated in Fig. 1. In general, deep learning models are particularly adept at capturing the broader meaning or semantic information in images, but they may not always excel in detailing finer aspects. Therefore, we use a data-driven deep convolution model to initially determine the complex amplitude of the sample. This information then assists the physics model in reconstructing, thereby enriching the process with added context and deeper semantic insights. The deep and physics models in our framework analyze data differently, with one focusing on the spatial domain and the other focusing on the frequency domain, each with its strengths. We utilize another deep model to leverage these differences, extracting and using paired features from both models. This approach effectively reduces noise from the physics model output, significantly enhancing the overall quality of the final output. Our method effectively merges the problem-solving capabilities of deep models with the generalizing strengths of physical models. It also eases the challenge for neural networks to generalize finer details, particularly in scenarios with limited sample data. In our tests, we effectively utilized a resolution target, showcasing the framework's effectiveness even with its data-driven components. This approach, challenging with previous methods, highlights our architecture's strengths. It underscores the potential applicability of our framework in fields like digital pathology, where deciphering the deeper meanings within images is essential.

This paper is organized as follows. Section 2.1 expounds upon the underlying principles of the multiplexed FPM which is programmable illumination. The details of our framework are discussed in Section 2.2, while Section 2.3 presents the principle of the data augmentation we conduct for rare samples. In Section 3, experiments are conducted to validate the capability of our framework to image with high reconstruction quality. Conclusions are subsequently synthesized in Section 4.

## 2 Method

In FPM, the reconstruction with limited number of LR images is a complex and challenging nonlinear optimization task. For these kind of problems, data-driven methods tend to produce overly smooth results that lack detail due to their generalization limits, despite containing significant semantic information. Meanwhile, physics models often risk getting stuck at local optima during the reconstruction phase, leading to unsatisfactory results. This issue becomes more pronounced when the number of LR images available is limited. To address this, our framework integrates two deep learning modules and a physics model module, encompassing a three-step process: end-to-end image reconstruction, reconstruction using the physics model, and enhancing image quality by leveraging paired features from both methodologies. We first reconstruct the HR image by using an end-to-end neural network. And then, the outcome is used to initialize the physics model's Fourier object layer, effectively infusing more semantic context into the model. Nevertheless, despite the introduction of additional information enhanced the physics model's imaging quality, the results still contain some periodic noise. To combat this issue, we've integrated another deep learning module into our framework. This module is tasked with harnessing and leveraging the paired features to refine the final image quality further. For instance, as shown in Fig. 2b, periodic noises like ringing and stripe noises that lack clear semantic information can be easily identified by comparing the outputs of the first step and the physics module. Our framework is designed with the flexibility to swap out the main components of both the deep learning modules and the physical model with alternative structures or models, if needed.

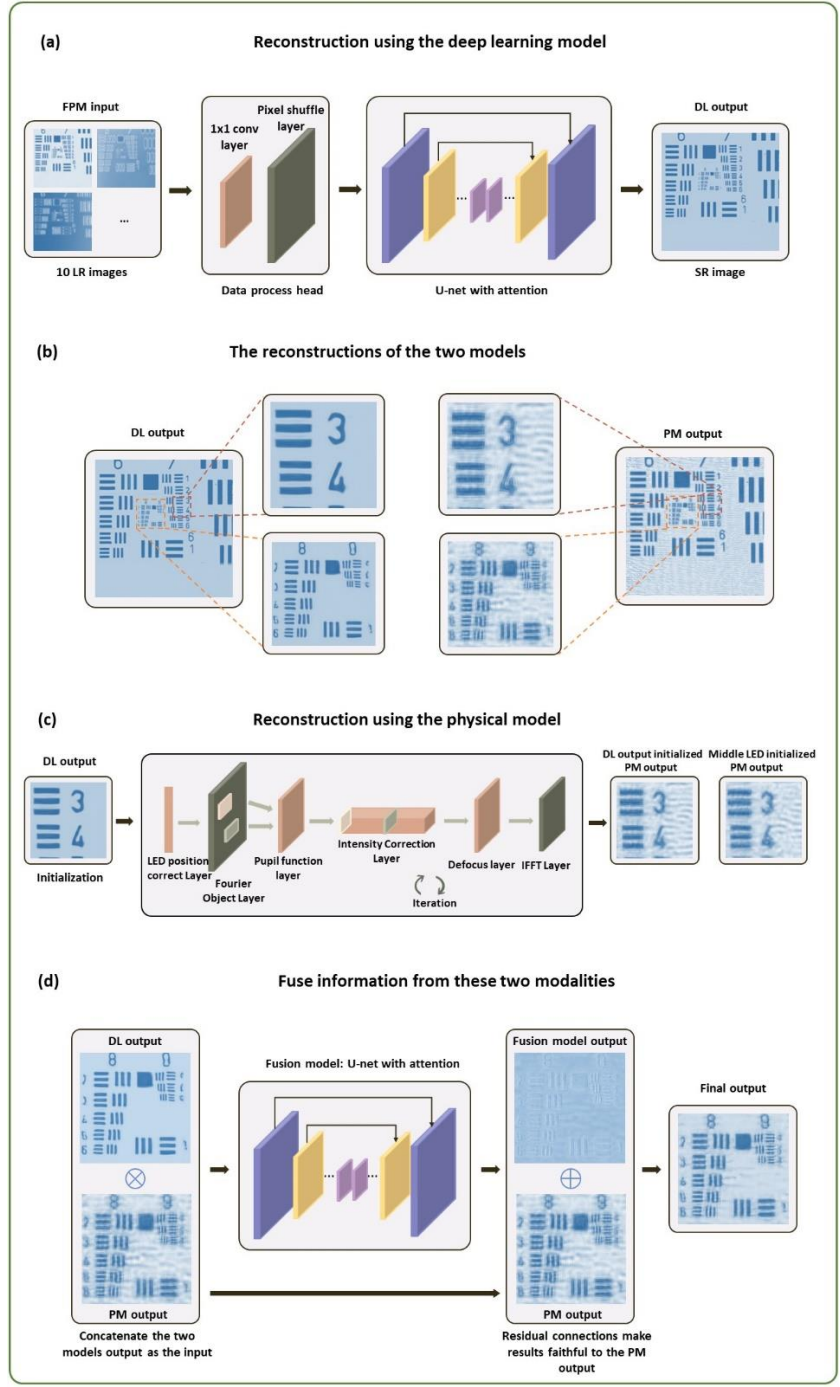

**Fig. 2.** The details of our framework. We describe the first step of the framework in (a), the second step in (c), and the last step in (d). We show the noise introduced by the physical model (PM) and an example output of the deep learning (DL) model in (b).

## 2.1 End-to-end image reconstruction model

We opted for a U-Net network with an attention mechanism [26] as the core of our deep learning modules. Created for cell-related semantic segmentation tasks [27], U-Net is known for its excellent performance and straightforward design characteristics. This has earned it widespread use in the biomedical realm [28], allowing it to address complex optical challenges [25, 32] effectively. Unlike most tasks related to natural or biomedical images, the FPM reconstruction task usually requires multiple images as input. As a result, we designed an extra head module to process these images. We merge the collected LR images along their channel dimension and resize them using a 1x1 convolution layer with a pixel shuffle layer to prepare them for processing in the first step, as shown in Fig. 2a.

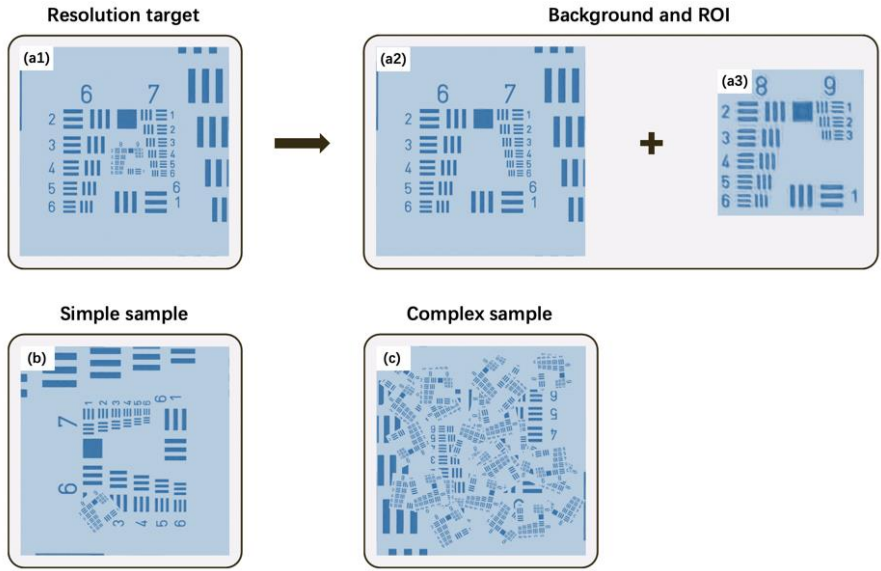

**Fig. 3.** The overview of our proposed data augmentation methods. (a1) The ground truth of the resolution target. (a2) The background we extract. (a3) The Region of Interest (ROI) of the resolution target. (b) An example of simple samples. (c) An example of complex samples.

It is well known that the quality of the training data set determines the performance of the neural network. However, when it comes to rare samples, collecting enough high-quality data is a significant challenge. For example, the resolution target is a typical representative of rare samples. The most informative part of it is limited and comprises just a small fraction of the entire sample. This often results in networks overlooking these crucial areas during training. Therefore, it is particularly important to collect large amounts of data for training. However, capturing data by taking multiple imaging processes of the same resolution target is labor-intensive and consumes substantial resources. In contrast, producing vast quantities of high-quality data through simulations that accurately model the imaging process proves to be a far superior approach. Nevertheless, blindly augmenting data without considering sample characteristics may not aid in training and could even be detrimental. It could also reduce the model's interpretability. To address this, we adopted a data augmentation method that considers sample characteristics. We start by separating the samples into the background and Region of Interest (ROI), as shown in Fig. 3 (a1), (a2), and (a3). Then, we mix these ROI parts back into the background in various rotated and flipped ways and at different spots. The corresponding LR images is obtained by performing a forward propagation process of the physics model. We categorize the samples based on how many detailed parts they have, labeling them as either simple or complex, as shown in Fig. 3 (b) and (c). Under the training process, our model learns to rebuild images from simple datasets first and then gets better at picking up details with more complex ones. Notably, with the framework we proposed, even if the neural network is only trained on simulation data, it can still accurately reconstruct HR images from actual, real-life experiments.

## 2.2 Physics model

In the framework of FPM, the sample under observation is assumed to be a two-dimensional flat thin layer. Its optical properties can be expressed by the complex transmission function $o(r) = \exp(i\varphi(r) - \mu(r))$, where $\varphi(r)$ represents the phase modulation distribution, and $\mu(r)$ indicates the absorption distribution. For a typical FPM light source, usually a programable LED array, illumination from each LED unit located at coordinates (x, y) is treated as a spatially coherent local plane wave. The outgoing wave passing through the sample can be described as $u(r) = o(r)\exp(ik_m \cdot r)$. In this equation, $k_m = (\sin\theta_{xm}/\lambda, \sin\theta_{ym}/\lambda)$ indicates the frequency shift relative to the sample's spectrum center, where '$\lambda$' denotes the wavelength, and $(\theta_{xm}, \theta_{ym})$ denote the angle of incident wave illumination, respectively. Compared with conventional FPM framework, in which one LED is lit up once, LED illumination multiplexing is proved to be an efficient way to increase the imaging speed of programable illumination computational microscopy [8, 12, 13]. When multiple LEDs at different positions are lit up simultaneously, the corresponding single-shot low-resolution image captured by the camera can be described as the sum of the intensities of the illumination from each individual LED, as demonstrated in Eq. (1). In this equation, '$I$' stands for the total intensity, $o(k - k_{mi})$ is used to describe the spectrum shift with different illumination angles, $P(k)$ represents the pupil function, $F$ represents the two-dimensional Fourier transformation and 'N' signifies the number of LEDs illuminated simultaneously. We adopt the neural network model introduced by Delong Yang et al. [32] to model the image process and reconstruct the HR image, as illustrated in Fig. 2c.

$$I_{multiple} = \sum_{i=1}^{n} |F(o(k - k_{mi})P(k))|^2 \qquad (1)$$

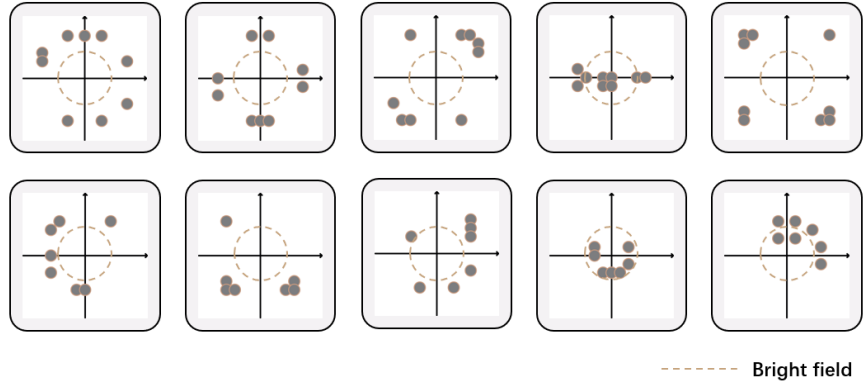

**Fig. 4.** The illumination model we generated. The gray circles represent the corresponding sample spectrum range when different LED lights are on.

Since the reconstruction process of FPM involves a nonlinear phase retrieval algorithm, it is impractical to explicitly obtain the optimal multiplexed illumination model in each observation. Therefore, in our previous work, we have designed an unsupervised adaptive illumination model generation method to get an optimal illumination multiplexing strategy to some extent. [8]. Using LR images taken when the central light is on as a prior, we generated an illumination model for multiplexed FPM. This model was then applied to all the image acquisition parts in our work. As shown in Fig. 4, the illumination model contains a total of 10 illumination patterns, which is much less than the sequential illumination model in conventional FPM framework.

*2.3 Deep-learning fusion model*

In the last step, we concatenate the outputs from both the deep learning and physics models as input for the fusion model. We also included a residual connection that merges the output from the physics model with the fusion model's results before final output to boost the framework's ability to generalize as shown in Fig. 2d. In the training phase of the deep learning model and the iteration of the physics model, Mean Absolute Error is employed as the loss function for both, as specified in Eq. (2). However, although their optimization goals are the same, they often produce different results, as shown in Fig. 2b. The reason for these varied outcomes lies in the fundamental structural variances between the neural network and the physics model, which lead them on distinct trajectories during the optimization process. Here, we can describe the collection process of the LR image as $LR = fp(HR)$, and the reconstruction processes of deep learning model and physics model as Eq. (3) and Eq. (4), where $fp(LR)$ represents the forward propagation process of FPM, $LR$ stands for the collected low-resolution image, and HR represents the true complex amplitude of the sample.

$$L_1(x,y) = \sum_{i=1}^{n} \frac{1}{n} |x_i - y_i| \tag{2}$$

$$SR_{dl} = f(LR) = f(fp(HR)) \tag{3}$$

$$SR_{pm} = g(LR) = g(fp(HR)) \tag{4}$$

Eq. (3) and (4) demonstrate how each reconstruction method transforms the original complex amplitude of the sample into two distinct feature spaces, thereby generating two different modalities. Therefore, any specific area of the sample can be described by the vector $(SR_{dl}, SR_{pm})$, which represents the paired features contained in the two modalities. It is important to note that extracting information from paired features presents a challenge when working with single modality data [31].

## 3 Experiments

This chapter focuses on the experimental setup and the results our framework has delivered. We conducted the training and inference of the network on an RTX3090 with a memory size of 24G. The superiority of our framework was qualitatively and quantitatively analyzed on the USAF chart and biological samples.

## 3.1 Experiments on USAF chart

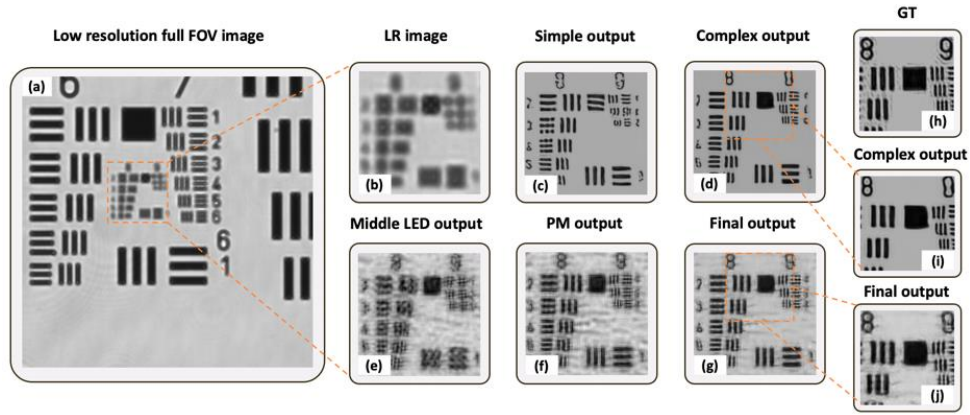

**Fig. 5** Experiments results on USAF chart. (a) The captured image illuminated by the middle LED. (b) The low-resolution area without reconstructing (c) The reconstruction result of deep learning (DL) model trained on simple dataset. (d) The reconstruction result of deep learning (DL) model trained on complex dataset. (e) The reconstruction result of physical model (PM) initialized by the image middle LED illuminated. (f) The reconstruction result of physical model (PM) initialized by the output of DL model trained on complex dataset. (g) The final reconstruction result of our framework. (h) The ground truth (GT) reconstructed from 121 low-resolution images captured sequentially. (i) The details of the deep learning output. (j) The details of the final output.

Following the method proposed in Section 2.3, we first created a complex dataset and a simple dataset for the training of the first deep learning module. Each dataset contains 768 paired data, divided into training, validation, and test sets in a 4:1:1 ratio. Each image pair included a HR image (1024x1024) and ten corresponding LR images (512x512). Due to memory constraints and the inherent workings of convolutional neural networks, we capped the network's output at a 256x256 size. Therefore, the obtained data was evenly divided into 16 equally sized small blocks, resulting in two datasets, each containing 768x16 groups of data. We first trained a randomly initialized network for 200 epochs on the simple dataset and then continued training for 200 epochs on the complex dataset. At this point, we obtained the end-to-end deep model for image reconstruction. Subsequently, we used the model trained on the simple dataset and the physics model to reconstruct the simple training dataset, thereby obtaining the dataset for training the fusion model. We avoided using the more advanced model trained on the complex dataset for this stage because its high performance could make the fusion model overly dependent on its results, which would impede the model's ability to learn and use paired features effectively.

To demonstrate that our proposed framework effectively combines the powerful modeling capabilities of deep learning models with the strong generalization ability of physical models, we used models trained on simulated data for actual experimental data reconstruction. Specifically, we kept the experimental parameters consistent with the simulation process, placed the USAF chart 97mm from the LED array, and used 470nm wavelength light with a 20nm bandwidth for illumination. We observed with a lens with a 4x magnification and a numerical aperture of 0.1NA and recorded LR images with a camera that boasts a dynamic range of 71.89 dB and a pixel size of 2.4 μm. As shown in Fig. 5b and 5c, the data augmentation method we used prompted the network to pay more attention to the areas of interest, achieving better reconstruction performance. Fig. 5d and 5f illustrate that with the deep learning model added semantic insights, the physics model achieved superior recovery results. In contrast with previous work of multiplexed FPM, our framework achieved high-quality Fourier ptychographic microscopy imaging with only 10 collected LR images as shown in Fig.5 g.

## 3.2 Experiments on Biological samples

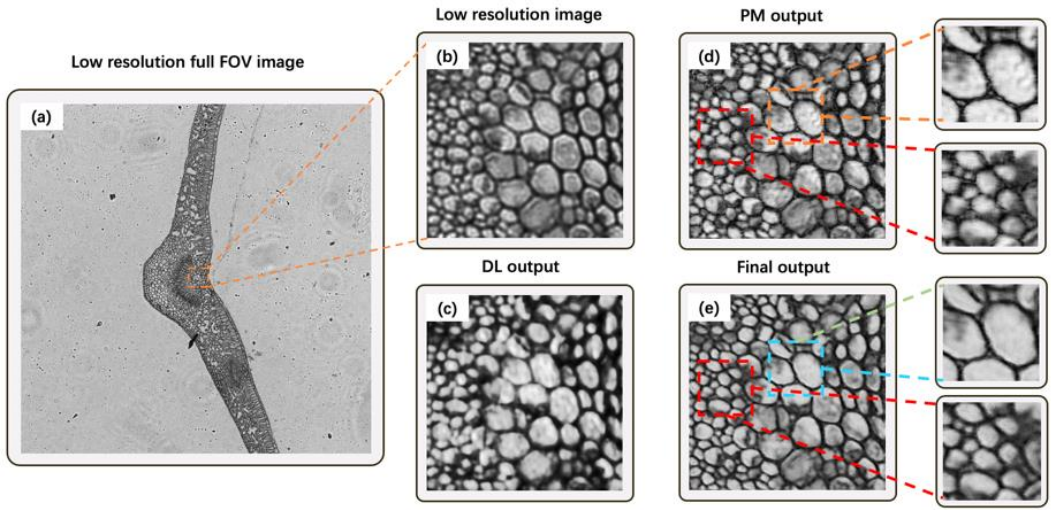

**Fig. 6** (a) The captured image illuminated by the middle LED. (b) The low-resolution area without reconstructing. (c) The reconstruction result of the first deep learning (DL) model. (d) The reconstruction result of physical model (PM). (e) The final reconstruction result of our framework.

We conducted tests with biological samples to showcase the real-world applicability and impressive generalization capabilities of our framework. For illumination, we used light with a wavelength of 470nm and a bandwidth of 20nm, and we captured LR images using a camera with a 6.5 μm pixel size. Our training dataset for the reconstruction network included various plant sections such as lotus stem, bamboo stem, and mint stem, while we used broad bean stem sections for the fusion network training, and privet leaf sections to evaluate the framework's performance. It is worth mentioning that we used completely different biological samples for training and testing, closely approximating real-world application scenarios. The ground truth images for the training set were created by reconstructing 121 LR images taken under various illumination angles with a physics model. The results, as seen in Fig. 6, illustrate that even with only 10 captured images, our framework could reconstruct the privet leaf section with remarkable clarity, underlining its robust generalization and imaging prowess.

### 3.3 Ablation experiments

In this section, we conducted comprehensive ablation experiments to confirm the effectiveness of our fusion model in leveraging paired features from two modalities and enhancing final image quality. We experimented with the broad bean slice data mentioned in Section 3.2 by changing the input combination of the fusion model during testing. And divided the whole image into 441 smaller segments, allocating them in a 7:3 ratio between training and testing sets (we skipped creating a validation set as it didn't require extra adjustments of hyperparameters). Furthermore, we excluded blank areas by sorting each cut image based on its contrast level to avoid extra interference.

The input combinations for model testing were set as (DL_output, DL_output), (PM_output, PM_output), and (DL_output, PM_output), with the results detailed in Table 1. When only the reconstruction results of the deep learning model were used as input, the final output of the fusion module was nearly consistent with the performance of the physics model. This occurs because without extra information, the model finds it challenging to accurately reconstruct the finer details, and as a result, it tends to adhere closely to the physical model's output. This adherence is due to the residual connections, as explained in Section 2.2. When only the reconstruction results of the physics model were used as input, the results improved. This is because, while learning pairwise features, the single-modal features are also learned by the network. The best results came when we combined inputs from both the deep learning and physics models. This clearly showed that our framework could identify and utilize the paired features from the results of different methods, leading to a more precise result.

When using deep learning for recovering blank regions in samples, we observed the appearance of concentric rectangular noise in the recovered areas. This pattern of noise is distinctly different from the global, uniform noise typically seen in results from the physics model, as depicted in Fig. 7. The concentric rectangular noise is due to the convolutional neural network's dependency on its depth to gradually expand the receptive field's boundaries, leading to noise accumulation at these edges. In contrast, the physics model operates in the frequency domain, providing a consistently global field of view across the image. As a result, the depth model and the physics model can be viewed as analyzing and interpreting the acquired data from different perspectives. The distinct noise patterns in the results from the two reconstruction methods align with the theory we discussed in Section 2.2. This observation from our experiments validates the presence of paired features, offering practical evidence of their existence. Through leveraging paired features, our proposed architecture effectively preserves detailed information while filtering out the noise observed in the results mentioned above as shown in Fig 7d.

**Table 1** Comparison of results from different methods in ablation experiments.

| Method | Evaluation metric(amplitude) | | | |
| --- | --- | --- | --- | --- |
| | SSIM ↑ | PSNR ↑ | NIQE ↓ | LPIPS ↓ |
| Reconstruction with DL model | 0.532 | 17.4 | 50.4 | 0.195 |
| Reconstruction with PM | 0.594 | 22.8 | 62.2 | 0.129 |
| Reconstruction with our framework (DL, DL) | 0.587 | 22.3 | 70.0 | 0.124 |
| Reconstruction with our framework (PM, PM) | 0.728 | 26.2 | 49.9 | 0.108 |
| **Reconstruction with our framework (DL, PM)** | **0.740** | **26.8** | **48.8** | **0.108** |

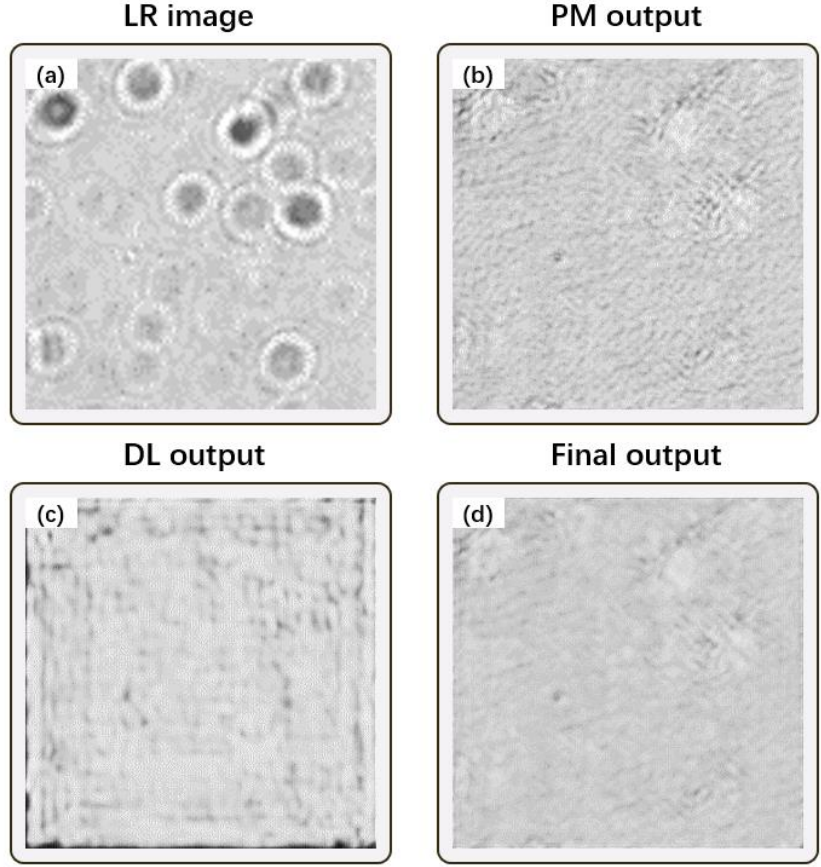

**Fig. 7** The comparison of the outputs between the deep learning and physics model. (a) The low-resolution area without reconstructing. (b) Reconstruction results of sample blank areas using physics model. (c) Reconstruction results of sample blank areas using data-driven deep learning model. (d) The final reconstruction result of our framework.

## 4  Conclusion and discussion

In FPM, physics models can yield high-quality reconstruction results for illumination multiplexing workflows. However, lighting up multiple LEDs in different locations at once complicates the reconstruction process for physics models. End-to-end deep learning methods are now increasingly used to solve optical inverse problems, especially those with limited captured LR images like multiplexed FPM. Nevertheless, these methods often face limitations in their ability to generalize, restricting their applicability in practical scenarios. In this paper, we proposed a hybridization framework that integrated the generalization strengths of physics models with the rapid reconstruction and complex information processing abilities of deep models. It also effectively utilized the paired features from different reconstruction methods to further enhance imaging quality. Furthermore, with the data enhancement method proposed, we qualitatively analyze the imaging capabilities of our framework in experiments using resolution target samples which is difficult for previous data-driven work. Our framework is also well-suited for areas with limited samples, like studying novel transparent materials or diagnosing rare diseases. To prove its wide applicability, we've successfully trained and tested it with different plant samples, a process that aligns closely with the real-world tasks of botanists. By analyzing the noise characteristics from different reconstruction methods, we were able to confirm the presence of paired features. This indicates that when physical constraints are not enough, results from varying imaging techniques can be combined by fusion module to effectively improve imaging accuracy.

In contrast to the DIP method, our framework minimizes additional optimization challenges while integrating extra constraints. This approach is particularly suitable for applications like whole slide imaging, which demand high temporal resolution, due to its strengths in high reconstruction speed and clear interpretability. While our framework has only been demonstrated in FPM with programmable illumination, its flexible design makes it easily adaptable to other imaging tasks. Additionally, it holds significant potential for addressing challenges in various computational imaging domains. In our future work, we aim to migrate this highly promising approach to popular research areas such as digital pathology diagnostics and tomography imaging.

*Acknowledgments*

National Natural Science Foundation of China (62275020).

**Caption List**

**Fig. 1** Here we show the imaging system we used and how our framework worked.

**Fig. 2** The details of our framework.

**Fig. 3** The overview of our proposed data augmentation methods.

**Fig. 4** The illumination model we generated.

**Fig. 5** Experiments results on USAF chart.

**Fig. 6** Experiments results on biological samples.

**Fig. 7** The comparison of the outputs between the deep learning and physics model.

**Table 1** Comparison of results from different methods in ablation experiments.